# Proper Motions Of VLBI Lenses,
# Inertial Frames and
# The Evolution of Peculiar Velocities

C.S. Kochanek

T.S. Kolatt

Harvard-Smithsonian Center for Astrophysics, MS-51

60 Garden Street

Cambridge MA 02138

M. Bartelmann

Max-Planck-Institut für Astrophysik

Karl-Schwarzschild-Strasse 1, D–85748 Garching, Germany

## ABSTRACT

Precise determinations of the image positions in quad gravitational lenses using VLBI can be used to measure the transverse velocity of the lens galaxy and the observer. The typical proper motions are $\mu$as yr$^{-1}$, so the time scale to measure the motion is ten years. By measuring the dipole of the proper motions in an ensemble of lenses we can set limits on the deviation of the inertial frame defined by the lenses from that defined by the CMB dipole and estimate the Hubble constant. The residual proper motions after subtracting the dipole probe the evolution of peculiar velocities with redshift and can be used to estimate the density parameter $\Omega_0$. For $N$ lenses, VLBI measurement accuracies of $\sigma_\theta$, and a baseline of $T$ years, we estimate that the 2–$\sigma$ limit on the rms peculiar velocity of the lens galaxies is $3100(\sigma_\theta/10\mu$as$)($yrs$/T)/N^{1/2}$ km s$^{-1}$, and that the time required for the 2–$\sigma$ limit to reach the level of the local rms peculiar velocity $v_{0,rms}$ is approximately $10N^{-1/2}(v_{0,rms}/600$ km s$^{-1})(\sigma_\theta/10\mu$as$)$ years. For a ten year baseline and $N = 10$ lenses we expect the 1–$\sigma$ limit on the misalignment with the CMB dipole to be $\Delta\theta = 20°$ or equivalently to obtain an upper limit of $\Delta H_0/H_0 < 0.34$.

*Subject headings:* gravitational lensing – cosmology – peculiar velocities – VLBI – astrometry

## 1. Introduction

Extragalactic proper motions would be a major new cosmological tool if they could be measured. For example, an ensemble of distant extragalactic sources defines a rest frame, so the average dipole of the proper motions measures our peculiar velocity independent of the microwave background dipole (Kogut et al. 1993) and local kinematic determinations from galaxies or clusters (e.g. Riess, Press & Kirshner 1995; Postman & Lauer 1995; and see Dekel 1995; Strauss & Willick 1995 for reviews). In standard models the directions of the two dipoles should agree exactly, and the magnitudes of the two dipoles should agree up to the uncertainties in the distances to the sources. If the proper motion dipole and the CMB dipole agree in direction, then the alignment falsifies super-horizon sources for the CMB dipole (Langlois & Piran 1995; Paczyński & Piran 1990). Proper motions also limit the rotation of the reference frame, where the rotation may reflect a globally rotating universe (e.g. Birch 1982). If we assume that the proper motion dipole, $\mu$, and the CMB dipole, $v_\odot$, agree in magnitude and direction, then we can determine $H_0$ by comparing the lens redshifts to their distance $H_0^{-1} \propto D \propto v_\odot/\mu$. The residual proper motions, after subtracting the dipole due to our motion, measure the peculiar velocities of galaxies at cosmological distances. By measuring the proper motions of higher redshift sources, we can begin to observe the evolution of proper motions with redshift, thereby having an independent test of the predictions of structure formation models and the density parameter $\Omega_0$.

Outside the Galaxy, only VLBI has sufficient angular resolution to detect proper motions because of the large distances. At a characteristic cosmological (proper motion) distance of $D = 1 D_{Gpc} h_{75}^{-1}$ Gpc ($1 h_{75}^{-1}$ Gpc corresponds to $z = 0.3$ for $\Omega_0 = 1$ and $H_0 = 75 h_{75}$ km s$^{-1}$ Mpc$^{-1}$) an object with a transverse velocity of $\omega = 10^2 \omega_{100}$ km s$^{-1}$ shows a proper motion of $\mu = 0.021 \omega_{100} D_{Gpc}^{-1}$ $\mu$as yr$^{-1}$. Since realizable VLBI accuracies are $\simeq 10$ $\mu$as at cm wavelengths (e.g. Marcaide, Elósegui & Shapiro 1994; Campbell et al. 1994; Campbell 1995), only velocities $\gtrsim 5000 D_{Gpc}$ km s$^{-1}$ are detectable in one decade. As a result, extragalactic proper motions can be measured only in nearby maser sources [e.g. the disk in NGC 4258 (Miyoshi et al. 1995) where $D = 6.4$ Mpc and $\omega \sim 600$ km s$^{-1}$] or in superluminal jet sources. Superluminal jets combine both high physical velocities ($v \sim c$) with relativistic aberration (see Zensus & Pearson 1987) to produce proper motions between the AGN core and the emitted "blobs" of order mas yr$^{-1}$.

Unfortunately, the peculiar and virial velocities of galaxies and clusters of $100$ km s$^{-1} \lesssim v \lesssim 1000$ km s$^{-1}$ are too small for direct detection given current life expectancies. If the standard error in the position of a source is $\sigma_\theta = 10\sigma_{\theta 10}$ $\mu$as, then the time scale for a $1-\sigma$ measurement is $670 \omega_{100}^{-1} D_{Gpc} \sigma_{\theta 10}$ years. Barring a dramatic increase in life expectancy, we need sources with two features missing in typical VLBI sources. First, we need an effect like relativistic aberration to magnify the intrinsic velocities by a factor of 10 to 100. Second, the sources must have nearby reference sources. Proper motions must be measured relative to another source, and the more distant the reference source becomes, the more difficult it is to control the systematic uncertainties in the relative positions.

There are VLBI sources that have these features – gravitational lenses. Lenses magnify the source, so the apparent velocity of the lensed images is boosted by the magnification tensor. Particularly in the quad lenses, the typical magnification of the images is several tens, exactly the magnitude needed to make the proper motions measurable on "experimental" time scales. The motions in lenses include not only that of the observer and the source, but also the lens galaxy. In fact, the motion of the lens galaxy probably dominates over the source motion. The lens galaxy lies at fairly low redshifts $0.1 \lesssim z \lesssim 0.5$ (some lens galaxies can be very close to the observer – the lowest is $z \simeq 0.031$ for the optical quad lens 2237+0305, Huchra et al. 1985), so they are both closer to the observer and have larger peculiar velocities than the sources. Finally, by measuring proper motions in multiply imaged sources, we can measure the relative positions of the images rather than their absolute positions. This eliminates the need for an external reference source, and minimizes the systematic errors because all the sources/images are confined to a small angular region. Moreover, by measuring the relative motions of the images, we typically double the apparent velocity because pairs of highly magnified images generally have anti-parallel motions. If we scale from existing radio lens surveys (MG, Lawrence et al. 1984; Burke 1990; JVAS, Patnaik et al. 1992a; and CLASS, Myers et al. 1995) we expect 20 flat spectrum lenses brighter than 100 mJy, 40 brighter than 50 mJy, and 300 brighter than 10 mJy on the sky, and the bright samples are evenly divided between two and four image systems. Proper motions in gravitational lenses have previously been discussed in the context of superluminal sources. Chitre & Narlikar (1979, 1980, also Chitre & Saslaw 1989) discuss gravitational lensing as a possible cause of superluminal motions, and Gopal-Krishna & Subramanian (1991, 1996) discuss it as a means of proving that compact double radio sources are gravitational lenses.

In §2 we discuss the CMB dipole, our peculiar velocity, and the dependence of peculiar velocities on redshift and cosmology. In §3 we discuss the magnification of proper motions by gravitational lenses. In §4 we discuss the current VLBI accuracies and the feasibility of measuring the magnified proper motions. In §5 we discuss using the mean proper motions of lensed VLBI sources as an inertial reference frame, and in §6 we discuss using the residual proper motions to study the evolution of peculiar velocities and discuss systematic biases in the measurements. In §7 we consider effects other than peculiar velocities that may contribute to the observed magnified proper motion. Finally, in §8 we summarize our results.

## 2. Peculiar Velocities

The advantage of peculiar velocities as tools for studying large scale structure is that galaxies are used only as "test particles" moving in the gravitational field, thereby avoiding the unsolved problems associated with galaxy formation and the type of matter inducing the acceleration. Existing techniques determine the solar system peculiar velocity from the CMB dipole anisotropy to be $v_\odot = 386.6$ km s$^{-1}$ (Kogut et al. 1993), and the peculiar velocity of the Local Group (e.g. Yahil, Tamann & Sandage 1977) to be $v_{LG} \simeq 600$ km s$^{-1}$.

The determination of peculiar velocities is strongly limited by the accuracy of distance indicators. A typical distance indicator has a fractional error $\epsilon$, where the value of $\epsilon$ ranges from 0.05 for Type Ia supernovae (Riess et al. 1995) to 0.21 for the D$_n$–$\sigma$ relation of elliptical galaxies (Dressler et al. 1987). On top of this intrinsic scatter in the correlation between the distance dependent and the distance independent quantities, there are also systematic errors that can only be removed statistically. For the measurement accuracy to be comparable to the radial peculiar velocity, $v_{rms}/\sqrt{3}$, we must observe

$$N \simeq 3 \left( \frac{\epsilon\, cz}{v_{rms}} \right)^2 \simeq 19 \left( \frac{\epsilon}{0.15} \right)^2 \left( \frac{cz}{10000 \text{ km s}^{-1}} \right)^2 \left( \frac{600 \text{ km s}^{-1}}{v_{rms}} \right)^2 \tag{1}$$

independent objects at redshift $z$. This exceeds the *local* number of IRAS galaxies in a $(5\, h_{75}^{-1} \text{Mpc})^3$ volume by a factor of $\sim 2.5$. Even at distances much smaller than 10000 km s$^{-1}$, the intrinsic scatter is large, and averages over large volumes ($\sim (10\, h_{75}^{-1} \text{Mpc})^3$) are used to reduce the noise level. As a result we measure only coarse averages of the peculiar velocity over many galaxies rather than the peculiar velocities of individual galaxies. We estimate the raw, unaveraged peculiar velocity from the residuals about the mean peculiar velocities corrected for the estimated intrinsic scatter in the Hubble diagram, or from the pair-wise velocities of galaxies. The best observational estimates for the rms peculiar velocity range from 430 km s$^{-1}$ to 700 km s$^{-1}$, although estimates can be as low as 300 km s$^{-1}$ and as high as 1800 km s$^{-1}$ depending on the sample and the analysis method (see Strauss, Cen & Ostriker 1993).

Structure formation theories predict the evolution of the rms peculiar velocity, and when coupled with the COBE normalization (Górski 1994, Górski et al. 1994, White & Bunn 1995) for the power spectrum, predict the amplitude of the rms peculiar velocity today. Linear theory (e.g., Peebles 1993, §21) predicts that

$$v_{0,rms}^2 = \frac{H_0^2\, f^2(\Omega_0, \Lambda_0)}{2\, \pi^2} \int_0^\infty P(k)\, dk, \tag{2}$$

where a zero subscript denotes a value determined at the current epoch, $H_0$ is the Hubble constant in units of km s$^{-1}$ Mpc$^{-1}$, $P(k)$ is the COBE normalized power spectrum, and the function $f$ (Lahav et al. 1991) is

$$f(\Omega_0, \Lambda_0) \simeq \Omega_0^{0.6} + \frac{\Lambda_0}{70}\left(1 + \frac{\Omega_0}{2}\right) \quad \text{and} \quad f(\Omega_0, \Lambda_0, z) \simeq \left[\frac{\Omega_0(1+z)^3}{\Omega_0(1+z)^3 - \Omega_{0k}(1+z)^2 + \Lambda_0}\right]^{0.6},\tag{3}$$

where the second equation neglects the second term of the first, and where $\Omega_k = 1 - \Omega - \Lambda$ is the "curvature density" of the model. For example, in standard COBE normalized CDM, the predicted rms peculiar velocity today is $v_{rms,0} = 760$ km s$^{-1}$ with a 100 kpc filter. The rms peculiar velocity evolves with redshift as

$$v_{rms}(z) = v_{0,rms}\frac{f(\Omega_0, \Lambda_0, z)}{f(\Omega_0, \Lambda_0)}\frac{1}{\sqrt{1+z}}\ ,\tag{4}$$

and the rms transverse velocity causing the proper motions is just $\omega_{rms} = (2/3)^{1/2}v_{rms}(z)$. We must choose whether to use the theoretical models to predict both the evolution and amplitude of the velocities (eqns. (2), (3) and (4)), or simply to evolve the locally observed peculiar velocities to high redshift (eqn. (3) and (4)).

The rms proper motion as seen by an observer in the CMB rest frame is $\mu_{\rm CMB} = \omega/D_{OS}$, where $D_{OS}$ is the proper motion distance between the observer and the source. We use distances defined by

$$D(\Omega_0, \Lambda_0, z_1, z_2) = \frac{c}{H_0}\frac{\sinn\left\{|\Omega_{0k}|^{1/2}\int_{z_1}^{z_2}\left[(1+z)^2(1 + \Omega_0 z) - z(2 + z)\Lambda_0\right]^{-1/2}\ dz\right\}}{|\Omega_{0k}|^{1/2}}\tag{5}$$

where the function $\sinn(x)$ is $\sinh(x)$ for an open universe ($\Omega_k > 0$), $\sin(x)$ for a closed universe ($\Omega_k < 0$) and simply $x$ for a flat ($\Omega_k = 0$) universe (see Carroll, Press & Turner 1992). When $z_1 = 0$ this is the proper motion distance to redshift $z_2$.[1]

Given the apparent angular velocity in the CMB frame, the effective proper motion of a galaxy at $\vec{z}$ due to the combined solar system CMB dipole and the source velocity $\vec{\omega}_s$ is given by

$$\vec{\mu}_e = \frac{\vec{v}_\odot - (\hat{z}\cdot\vec{v}_\odot)\hat{z}}{D_{OS}} - \frac{\vec{\omega}_s}{D_{OS}}.\tag{6}$$

If, however, we are interested in the *relative* proper motion of a lens galaxy, L, at redshift $z_l$, and a source galaxy, S, at redshift $z_s > z_l$, as observed by an observer, O, then the effective

---

[1] For $z_1 \neq 0$ it is *not* the proper motion distance of an object seen at $z_2$ by an observer at $z_1$ – it lacks a redshift factor. The distances have the convenient algebraic property for a flat universe that $D_{13} = D_{12} + D_{23}$.

relative proper motion becomes

$$\vec{\mu}_e = \frac{\vec{v}_\odot - (\hat{z} \cdot \vec{v}_\odot)\hat{z}}{D_{OL}} \frac{D_{LS}}{D_{OS}} - \frac{1}{D_{OL}} \left( \vec{\omega}_l - \frac{D_{OL}}{D_{OS}} \vec{\omega}_s \right) \tag{7}$$

(Kayser, Refsdal & Stabell 1986, Miralda-Escudé 1991) where $\vec{\omega}_l$ and $\vec{\omega}_s$ are the transverse velocities of the lens and the source. For $z_l \le 0.3$ and $z_s \simeq 2$ we obtain in the range $0.1 \le \Omega_0 \le 1$

$$\frac{\omega_s}{\omega_l} \frac{D_{OL}}{D_{OS}} \le 0.20 \tag{8}$$

so we can usually neglect the source galaxy motion in eq. (7) and simply use the proper motion given by

$$\vec{\mu}_e = \frac{1}{D_{OL}} \left\{ \left[ \vec{v}_\odot - (\hat{z} \cdot \vec{v}_\odot)\hat{z} \right] \frac{D_{LS}}{D_{OS}} - \vec{\omega}_l \right\} . \tag{9}$$

This approximation, however, is not valid in the presence of large internal motions within the source (see §7). As discussed in the introduction, these proper motions are of the order of $\sim 0.1$ $\mu$as yr$^{-1}$ and thus can be observed only if appreciably magnified.

## 3. Lens Magnification of Proper Motions

In this section we first explore simple, generic models for the expected magnification of proper motions for fold caustics, cusp caustics, and symmetric lenses, and then we examine simple models for several of the known VLBI lenses. The general lens equation is

$$\vec{u} = \vec{x} - \nabla \psi, \tag{10}$$

where $\vec{x}$ and $\vec{u}$ are the angular coordinates of the lens plane and the source plane respectively, and $\psi$ is the two-dimensional effective potential of the lensing mass distribution (see Schneider, Ehlers & Falco 1992). The inverse magnification tensor is the Jacobian of the lens mapping

$$M_{ij}^{-1} = \delta_{ij} - \psi_{ij} ; \quad \psi_{ij} = \frac{\partial^2 \psi}{\partial x_i \partial x_j} , \tag{11}$$

and the magnification diverges on the critical curves where the determinant of the Jacobian is zero, $M^{-1} = |M_{ij}^{-1}| = 0$. Taking $\vec{\mu}_e$ of eq. (7) as the effective proper motion of the lens, the proper motion $\vec{\mu}_I$ of an image located at $\vec{x}_I$ becomes

$$\vec{\mu}_I = \overline{\overline{M}}(\vec{x}_I) \, \vec{\mu}_e \tag{12}$$

where $\overline{\overline{M}}$ is the magnification tensor with components $M_{ij}$.

When we need a concrete example, we use a singular isothermal sphere in an external shear field, for which

$$\psi = br - \frac{1}{2}\gamma r^2 \cos 2(\theta - \theta_\gamma), \tag{13}$$

where $b = 4\pi(\sigma/c)^2 D_{LS}/D_{OS}$ is determined by the velocity dispersion of the dark matter $\sigma$ ($\sigma \simeq 220$ km s$^{-1}$ for an $L_*$ galaxy), and the shear is characterized by a dimensionless strength $\gamma$ (typically $0.05 < \gamma < 0.15$) and an angle $\theta_\gamma$. This potential is a reasonable, but not perfect, model for lens statistics and lens models (see Kochanek 1991, 1996).

### 3.1. Motions In Particular Lenses

A source on one side of a fold caustic has two images on either side of the critical curve separated by

$$|\Delta x| = 2 \left[ \frac{2 \operatorname{tr}\left(\overline{\overline{M}}^{-1}\right)\Delta u}{|\vec{e} \cdot \nabla M^{-1}|} \right]^{1/2}, \tag{14}$$

where $\nabla M^{-1}$ is the gradient of the inverse magnification on the critical curve, $\vec{e}$ is the singular eigendirection of $\overline{\overline{M}}^{-1}$ on the critical curve, and $\Delta u$ is the distance of the source from the fold caustic. Two images are present for $\Delta u > 0$, and none are present for $\Delta u < 0$. To first order in $\Delta u$, only the component of the source velocity perpendicular to the caustic ($du_\perp/dt$) produces changes in the image separations. For the simple model potential (13), the time derivative of the distance between the images is

$$\left|\frac{d\Delta x}{dt}\right| = \frac{4b}{\Delta x} \frac{1}{3\gamma \sin 2\theta} \left|\frac{du_\perp}{dt}\right| \left(1 + \gamma \cos 2\theta + O(\gamma^2)\right) \tag{15}$$

to first order in the shear $\gamma$ for the model potential (13). Ignoring the small multiplicative corrections for the shear, the perpendicular velocity is magnified by $(4b)/(3\Delta x\gamma \sin 2\theta)$. The fold approximation is invalid for $\sin 2\theta = 0$, where the caustic has cusps.

The other generic feature of lens caustics is the cusp, where three rather than two images merge. We assume that the cusp is axisymmetric with respect to the $u_1$ axis, that its tip is at the coordinate origin, and that it points towards $u_1 \to \infty$. A source on the axis with $u_1 < 0$ has three images. One image is formed on the $x_1$ axis and has negative parity, the two other images have positive parity and are located symmetrically on opposite sides of the $x_1$ axis. The relative motion of the two images with positive parity for a source on the symmetry axis is (e.g. Schneider, Ehlers, & Falco 1992)

$$\left|\frac{d|\Delta x|}{dt}\right| = \left|\frac{4\psi_{122}}{\psi_{122}^2 + \frac{1}{3}\psi_{2222}(1 - \psi_{11})}\right| \left|\frac{1}{|\Delta x|}\right| \left|\frac{du_1}{dt}\right|. \tag{16}$$

This can be evaluated for any lens model, but we can find a more general solution for our simple potential (eq. (13)) that includes this case as a limit by explicitly solving for the image positions when the source is on the $u_1$ axis for $\theta_\gamma = 0$. If $|u_1| < b$ then the lens produces at least two images located at $\vec{x}_1 = \{(u_1 + b)/(1 - \gamma), 0\}$ and $\vec{x}_2 = \{(u_1 - b)/(1 - \gamma), 0\}$. If $|u_1| < u_c = 2\gamma b/(1 + \gamma)$ the lens equation has two additional solutions at $\vec{x}_3 = \{-u/2\gamma, y_c\}$ and $\vec{x}_4 = \{-u/2\gamma, -y_c\}$ where $y_c = b(1 - u^2/u_c^2)^{1/2}/(1 + \gamma)$. This case is illustrated in Figure 1. The relative velocity of images 3 and 4 is

$$\frac{d\Delta x_{34}}{dt} = \frac{1}{(1 + \gamma)} \frac{2}{\gamma} \frac{b}{\Delta x_{34}} \left[ 1 - \left( \frac{(1 + \gamma)\Delta x_{34}}{2b} \right)^2 \right]^{1/2} \left| \frac{du_1}{dt} \right| \tag{17}$$

where the cusp limit is $\Delta x_{34} \ll b$. The factor under the square root suppresses the relative motion of images 3 and 4 in the limit of a symmetric quad lens ($\Delta x_{34} \simeq 2b/(1 + \gamma)$, $u \ll u_c$), where the source is near the origin and the images are arranged in a cross with images 3 and 4 forming the top and bottom images (as in 2237+0305.) If we look at the relative velocity of images 2 and 3 for $u \ll u_c$, then

$$\frac{d\Delta x_{23}}{dt} = \frac{1}{2\gamma} \left| \frac{du_1}{dt} \right|, \tag{18}$$

so that for the symmetric lenses there is a large magnification of the relative motions of the images on different axes. Comparing these equations, and ignoring corrections of order $\gamma$, the typical magnification of proper motions in a four image lens is of order $b/2\gamma\Delta x \gtrsim \gamma^{-1}$. Figure 1 shows a time sequence for a typical lens at $z_l = 0.3$ moving at 500 km s$^{-1}$ along the symmetry axis of the potential $\psi$ defined by $b = 1.0$ arcsec and $\gamma = 0.1$. Notice the very rapid motions of images 3 and 4 compared to images 1 and 2. In the four-image lenses we would measure the highly magnified relative motions of 3 and 4, while in the two image systems we can only measure the much slower motions of images 1 and 2. [2]

To illustrate these analytic estimates, we took the observational data for four VLBI lenses (MG 0414+0534, Hewitt et al. 1992; B 1422+231, Patnaik et al. 1992b; and CLASS 1608+656, Myers et al. 1995), and fit them with our simple potential model. We do not get perfect fits to both the relative positions and fluxes of the images, but we do get reasonable quantitative estimates of the sensitivity of the lenses to proper motions. From the fit we determine the magnification tensor at the position of image $I$, $\overline{\overline{M}}(\vec{x}_I)$. If the effective source position relative to the lens shifts by $\vec{u}$, then the image position shifts by $\delta\vec{x}_I = \overline{\overline{M}}(\vec{x}_I)\delta\vec{u}$. We can measure either the relative motions of the images $\delta\vec{x}_I - \delta\vec{x}_J$ or the absolute motions of the images $\delta\vec{x}_I$. Figure 2 shows the polar diagrams of motion sensitivity for the three source plane                                    image plane known four-image VLBI lenses, and the classic double lens 0957+561. We show both the maximum absolute motion of any single image, and the maximum relative motion of any image pairs as a function of the direction of the effective velocity. Note the different angular scales in the different panels.

---

[2]In fact, in this model, images 1 and 2 show no relative proper motions!

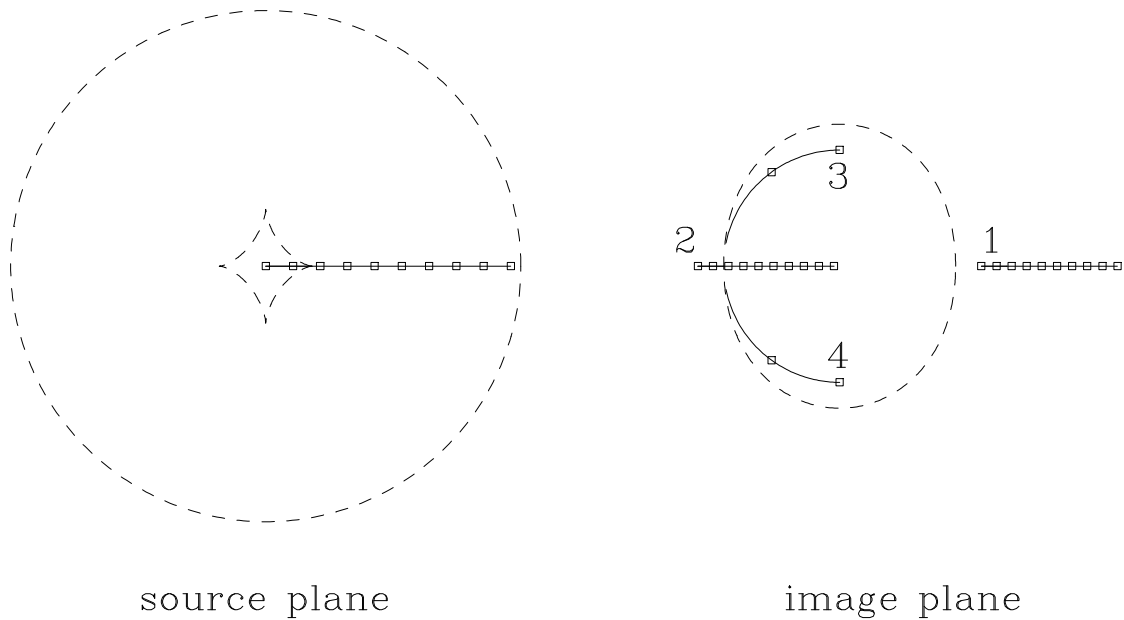

Fig. 1.— Proper motions. The left diagram shows the source plane, and the right shows the lens plane. The scale of the source plane is magnified by a factor of two. The dashed lines in the source plane show the outer radial pseudo-caustic and the inner astroid caustic, and the dashed line in the image plane shows the tangential critical line. The source and image motions are shown by the solid lines, and the points are spaced at time intervals of $10^6$ years for a lens at $z = 0.3$ moving at 500 km s$^{-1}$ and a lens critical radius of $b = 1.0$ arcsec. It takes $9.6 \times 10^6$ years to cross the critical radius. The images start at the image number labels for a source at the origin.

## 3.2. Statistics of Proper Motion Magnification

For the statistical analyses of §5 and §6 we need expressions for the mean square magnification of proper motions. In the isothermal model (eqn. (13)) the eigenvalues of the magnification tensor are effectively $\lambda_1 = 1$ and $\lambda_2 = M_I$ where $M_I$ is the magnification of image $I$, so if we know the magnification probability distribution, we can estimate the average magnification of proper motions. If we assume that the total magnification $M \gg 1$, then we can allow $\lambda_1 = 0$. For the four image lenses, the mean total magnification of all four images is $\langle M \rangle = C/\gamma$ where $C = 3.6$, and the integrated probability distribution for the total magnification is well approximated by $P(> M) = (\langle M \rangle / 2M)^2$ with $\langle M \rangle / 2 < M < \infty$.

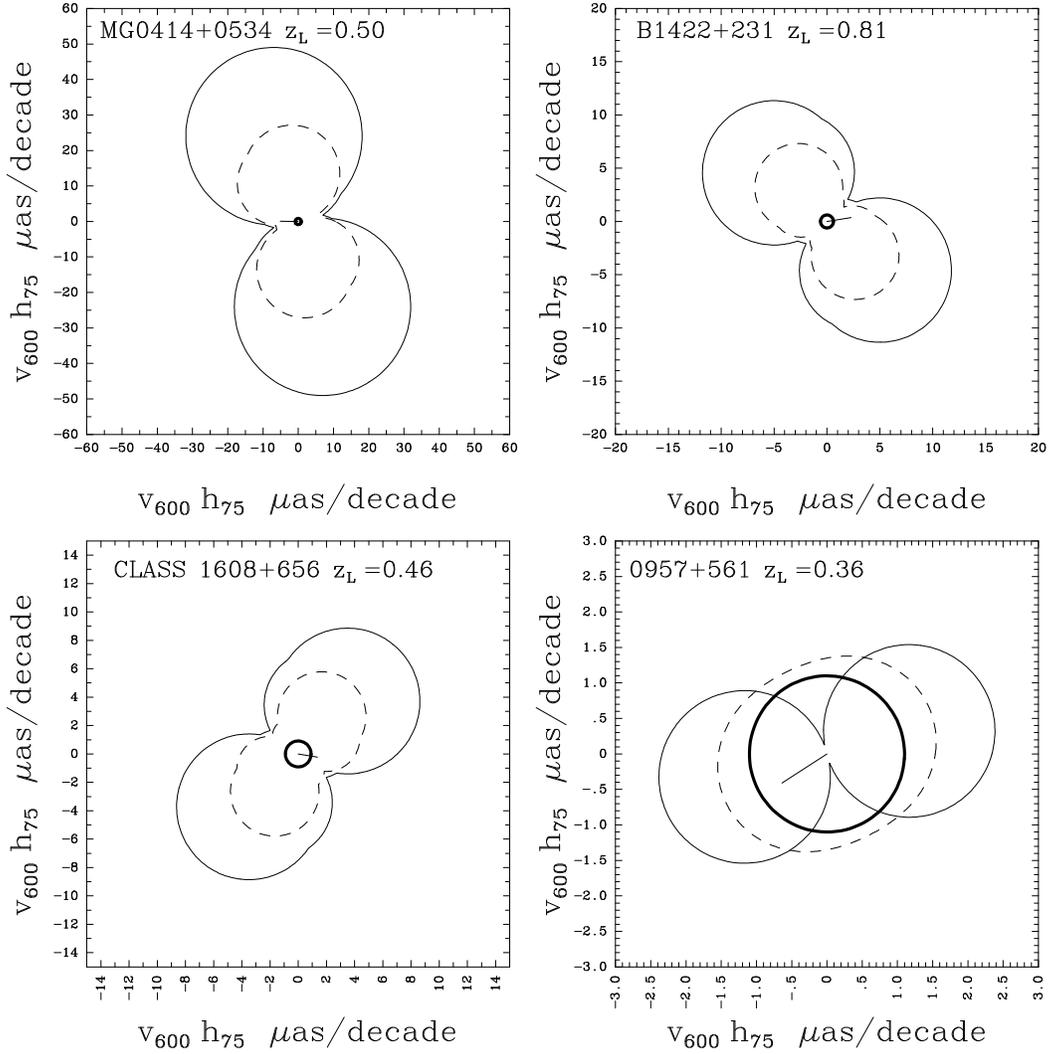

Fig. 2.— Polar sensitivity plots for proper motions in MG 0414+0534, B 1422+231, CLASS 1608+656, and 0957+561. The angle is the direction of the effective source velocity, and the magnitude is the resulting proper motion scaled to an effective velocity of $600 h_{75}$ km s$^{-1}$ and the listed lens redshift. The solid curve shows the maximum relative proper motion between the images, and the dashed curve shows the maximum absolute proper motion of any image. The solid vectors show the direction and magnitude of the effective velocity and proper motions corresponding to the projection of the CMB dipole velocity on the lens (north to top, east to left). The heavy solid circles show the unmagnified proper motions expected for a 600 km s$^{-1}$ velocity at the lens redshift. Note the different angular scales in the different panels.

For high magnifications, the image proper motions are dominated by folds, so the effective magnification of the proper motions is $M_\mu \simeq M$ because two images dominate the magnification with $M_a \simeq M_b \simeq M/2$ and $M_c \simeq M_d \ll M$ and the magnification tensors of the two highly magnified images will be nearly aligned. Thus the mean magnification of the proper motions is

$$\langle M_\mu \rangle = \int_{\langle M \rangle /2}^{\infty} dM \frac{dP}{dM} M = \frac{C}{\gamma}, \tag{19}$$

and the mean square magnification is

$$\langle M_\mu^2 \rangle = \int_{\langle M \rangle /2}^{M_{max}} dM \frac{dP}{dM} M^2 = \langle M_\mu \rangle^2 \ln \left[ \frac{2b}{\theta_s \langle M \rangle} \right]^{1/2}. \tag{20}$$

The latter integral is logarithmically divergent in the maximum magnification cutoff of $M_{max} \sim b/\theta_s \sim 10^3 (\text{mas}/\theta_s)$ where $b$ is the critical radius and $\theta_s$ is the source size. For statistical calculations we will use the approximation that the mean square value of the magnification eigenvalues that enters the statistical calculations is

$$\langle \lambda_1^2 + \lambda_2^2 \rangle \cong \langle M_\mu \rangle^2 = \frac{C^2}{\gamma^2}. \tag{21}$$

This is an *underestimate* of the true proper motion magnification because we neglect the logarithmic divergence, multiple image pairs, and magnification bias. Magnification bias makes flux-limited samples have higher mean magnifications than the mean of the magnification probability distribution.

## 4. VLBI Measurements

The observational issue in this project is the accuracy with which image positions can be determined using VLBI. We define $\sigma_\theta$ to be the $1-\sigma$ error in one coordinate component, and scale all estimates to $\sigma_\theta = 10\ \mu$as. The simplest measurement of the proper motions consists of two epochs separated by time $T$ with rms position errors $\sigma_\theta$, and this leads to rms measurement errors for a component of the relative proper motion of one image pair of

$$\langle e_l^2 \rangle^{1/2} \simeq \sqrt{2} \frac{\sigma_\theta}{T} = 14 \left( \frac{\sigma_\theta}{10 \mu \text{as}} \right) \left( \frac{\text{years}}{T} \right) \mu \text{as yr}^{-1}. \tag{22}$$

Additional measurements at intermediate epochs, and averaging over the independent image baselines will reduce the uncertainties, but we scale our estimates using this simple measurement model. The estimates we make in §5 and §6 will overestimate the uncertainties for any given value of $\sigma_\theta$.

Few astrometric VLBI observations closely resemble the problem of determining relative proper motions in gravitational lenses, because in most astrometric studies the errors are dominated by effects arising from the separation of the source and the nearest reference source. For example, Guirado et al. (1995) estimated the separation of 1928+738 and 2007+777 (separated by 5 degrees) to an rms accuracy of 200–300 $\mu$as, where the uncertainties were dominated by the tropospheric model and the position of the reference source. Marcaide, Elósegui & Shapiro (1994) estimated the relative proper motion of 1038+528 A and B (separated by 33 arcseconds). In this case both sources can be observed simultaneously, as is the case for the lens sources. Such sources can be "phase connected" and the possible precision increases greatly. The statistical uncertainties in the source positions were $\sim 2$ $\mu$as at 3.6 cm and $\sim 10$ $\mu$as at 13 cm, leading to a nominal proper motion of $31 \pm 2$ $\mu$as yr$^{-1}$ between 1981 and 1983. Unfortunately, an analysis of systematic errors in the measurement suggests that the true errors are nearly ten times larger, and their final estimate of the uncertainty in the proper motion was $\pm 22$ $\mu$as yr$^{-1}$ at 3.6 cm. Only one lens, 0957+561, has accurate, multiple epoch, astrometric data (Campbell 1995, Campbell et al. 1994). In this system (see §3) there are two images separated by 6 arcseconds, so it is even easier than in the case of 1038+528 A/B to determine the relative positions. At 6 cm the typical statistical uncertainty in the relative positions was 13 $\mu$as for each coordinate, but the actual scatter in the data (including earlier results by Gorenstein et al. 1984, 1988) was two to three times larger.

If we reject the possibility that the apparent motions in 1038+528 A/B and 0957+561 are physical (they correspond to motions with $v \gtrsim c$, but see §7), then current measurement accuracies are of order 20–30 $\mu$as. Fortunately, the quad lens systems are probably less sensitive to systematic errors than either of these systems, because the image separations are smaller (2 arcsec or less). This reduces the magnitude of systematic errors that are first (second) order in the separations by another factor of 3 (9) relative to 0957+561. The presence of four images allows the systematic errors to be estimated from even two epochs of data, because the best fit proper motion specifies only two of six relative motions. The residuals from the best fit proper motion are a combination of the intrinsic errors in the VLBI measurements and errors in the magnification tensor estimates. Space-VLBI and the use of higher observing frequencies may also reduce the uncertainties by improving the intrinsic resolution of the observations.

## 5.  The Mean Proper Motion and The Inertial Reference Frame

For each lens we measure the changes in the relative image positions over a time baseline $T$. In a lens with $n = 4$ images, there are $n(n-1)/2 = 6$ independent relative positions. The change in the relative positions of images $I$ and $J$ is $T\overrightarrow{\Delta}\mu_{IJ}$, where the relative proper motion is $\overrightarrow{\Delta}\mu_{IJ} = \vec{\mu}_I - \vec{\mu}_J$. We assume that the magnification tensor at each image $\overline{\overline{M}}(\vec{x})$ is exactly determined by a lens model, allowing us to solve for the effective proper motion $\vec{\mu}_e^l$ in each of $l = 1 \cdots N$ lenses.

To estimate the expected signal-to-noise ratio for detecting the proper motions we consider the case of a single image pair. The magnification of the relative proper motion is determined by the eigenvalues $\lambda_1$ and $\lambda_2$ of the difference in their magnification tensors $(\overline{\overline{M}}_I - \overline{\overline{M}}_J)$, and the relative orientations of the magnification eigenvectors and the peculiar velocity. If we average over the orientation of the lens, the location of the lens, the position of the source, the lens peculiar velocities, and use the ensemble average $\langle \lambda_1^2 + \lambda_2^2 \rangle = C^2 \gamma^{-2}$ of the proper motion magnification (see §3.2) , then the typical signal-to-noise ratio for a single lens is

$$\left\langle \frac{|\overline{\overline{M}}\vec{\mu}_e|^2}{e_l^2} \right\rangle^{1/2} = \frac{\langle |\overline{\overline{M}}\vec{\mu}_e|^2 \rangle^{1/2}}{\langle e_l^2 \rangle^{1/2}} = \frac{\langle \lambda_1^2 + \lambda_2^2 \rangle^{1/2}}{\sqrt{6}} \frac{v_\odot T}{D_{OL}\sigma_\theta} \left[ \frac{D_{LS}^2}{D_{OS}^2} + \frac{3}{2}\frac{\langle \omega_l^2 \rangle}{v_\odot^2} + \frac{3}{2}\frac{\langle \omega_s^2 \rangle}{v_\odot^2} \frac{D_{OL}^2}{D_{OS}^2} \right]^{1/2} \quad (23)$$

$$= \frac{0.0031}{\gamma} \frac{v_\odot}{100 \text{ km s}^{-1}} \frac{T}{\text{yrs}} \frac{\text{Gpc}}{D_{OL}} \frac{10\mu\text{as}}{\sigma_\theta} \left[ \frac{D_{LS}^2}{D_{OS}^2} + \frac{v_{rms}^2(z_l)}{v_\odot^2} + \frac{v_{rms}^2(z_s)}{v_\odot^2} \frac{D_{OL}^2}{D_{OS}^2} \right]^{1/2} . \quad (24)$$

In an $\Omega_0 = 1$ universe with $\gamma = 0.075$, $z_l = 0.3$, $z_s = 2.0$, $v_{rms,0} = 600$ km s$^{-1}$, and $v_\odot = 369$ km s$^{-1}$ we obtain an rms signal to noise ratio of $\simeq 0.25 h_{75}(10\mu\text{as}/\sigma_\theta)(T/\text{yrs})$. A temporal baseline of $T = 12 h_{75}^{-1}(\sigma_\theta/10\mu\text{as})$ years leads to a $3-\sigma$ detection for a typical quad lens.

In each lens the proper motion is the sum of a proper motion dipole produced by our motion relative to the inertial frame and the randomly oriented proper motions of the lens and the source. We can determine our proper motion by fitting a model proper motion dipole $\vec{\mu}_{em}^l(\vec{z}_l, \vec{A})$ to the measured proper motions in an ensemble of gravitational lenses, where $\vec{A} = (\Omega_0, \Lambda_0, H_0, v_{0,rms}, P(k))$ contains all the cosmological parameters. Formally, we would fit the data using a $\chi^2$ statistic

$$\chi^2 = \sum_{l=1}^N \frac{(\vec{\mu}_e^l - \vec{\mu}_{em}^l)^2}{\tilde{e}_l^2 + \mu_{pec\,m}^2(z_l; \vec{A})} . \quad (25)$$

The additional term of $\mu_{pec\,m}^2$ in the denominator includes the noise in determining the mean dipole from the rms proper motions of the lens, and the $\tilde{e}_l$ are the measurement uncertainties translated to the uncertainties in the unmagnified proper motions ($\tilde{e}_l = e_l / \langle |\overline{\overline{M}}| \rangle$). Neglecting the noise from the lens proper motions, the mean square value of the $\chi^2$ is simply $N$ times the first term of equation (23).

Two simple results are the uncertainties in estimating the direction of the dipole velocity and the value of the Hubble constant assuming that the CMB dipole and lens dipole are identical. If we define $\Delta\theta$ as the uncertainty in the angle between the true dipole velocity and the model dipole velocity, and $\Delta H_0 / H_0$ as the fractional uncertainty in the value of the Hubble constant, then the $1-\sigma$ limits ($\Delta\chi^2 = 1$) for $N$ lenses with the noise dominated by the measurement error are approximately

$$\frac{\Delta H_0}{H_0} \quad \text{or} \quad \sin\Delta\theta = 1.1 \left(\frac{\Delta\chi^2}{N}\right)^{1/2} \left(\frac{\gamma}{0.1}\right) \left(\frac{\sigma_\theta}{10\mu as}\right) \left(\frac{D_{OS}}{4h_{75}^{-1}\mathrm{Gpc}}\right) \left(\frac{10\mathrm{yrs}}{T}\right), \quad (26)$$

where $\langle D_{LS}^2 / D_{OL}^2 D_{OS}^2 \rangle = 6/D_{OS}^2$ is an average over the lensing cross sections. For $N = 10$ lenses the $1-\sigma$ limit is $\Delta\theta = 20°$ and the $2-\sigma$ limit is $\Delta\theta = 43°$ after ten years. This estimate is a *lower limit* for a given value of $\sigma_\theta$, because the uncertainties for a particular lens can be reduced by considering all $n(n-1)/2$ relative image positions and by making more than two measurements of the image positions, and furthermore we are underestimating the typical magnifications. Longer observing time scales do not help because the error is eventually dominated by the real proper motions of the lenses rather than observational uncertainties. The ultimate limit on the uncertainty of the proper motion dipole is constrained by the total number of lenses. As can be seen in Figure 2, the proper motion due to the peculiar velocity of the lens is typically larger than the proper motion due to the peculiar velocity of the observer.

Figure 3 shows the estimated dipole direction using the location of known VLBI lenses. We drew the peculiar velocities from a Gaussian distribution with $v_{rms,0} = 600$ km s$^{-1}$ scaled to $z_l = 0.3$ in an $\Omega = 1$ universe, used average magnification of $C/\gamma\sqrt{2} = 34$ with $C = 3.6$ and $\gamma = 0.075$, and Gaussian error distribution with $\sigma_\theta = 10$ $\mu$as yr$^{-1}$ over a time baseline of 10 years. We ran 50 Monte-Carlo realizations and determined the mean dipole direction and its dispersion. We also show, on the same plot, the inertial frame dipole as obtained if all lenses rotate about the Galactic pole with rotation velocity equal to their expected *rms* two-dimensional peculiar velocity (thinner cross).

## 6. Peculiar Velocities at Cosmological Distances

Once the observer's peculiar motion is subtracted, the remaining motions are due to the peculiar velocities of the lens and of the source. For an $\Omega_0 = 1$ universe, the peculiar velocities scale with redshift as $\langle \omega_l^2(z) \rangle = \omega_{0,rms}^2/(1 + z)$ where $\omega_{0,rms}^2 = (2/3)v_{0,rms}^2$ (see eq. (3)), so the lens cross section weighted average of $\langle \omega_l^2/D_{OL}^2 \rangle = 10\omega_{0,rms}^2 g^2(z)/D_{OS}^2$ where $g(z) \simeq (1 + z_s)^{1/9}$ for $\Omega_0 = 1$. The contribution of the peculiar velocity of the source galaxy is negligible compared to that of the lens (see eq. (8)), because the relative contributions are

$$\frac{\langle \omega_s^2 D_{OS}^{-2} \rangle}{\langle \omega_l^2 D_{OL}^{-2} \rangle} = 0.1(1 + z_s)^{-11/9} \tag{27}$$

with the correct cross section weighted averages of the velocities. Thus the upper limit on the mean square peculiar motions scales as

$$v_{0,rms} < 3100 \left[ \frac{\Delta \chi^2}{N} \right]^{1/2} \left[ \frac{\gamma}{0.075} \right] \left[ \frac{\sigma_\theta}{10\mu as} \right] \left[ \frac{D_{OS}}{4g(z_s)\ \mathrm{Gpc}} \right] \left[ \frac{\mathrm{yrs}}{T} \right]\ \mathrm{km\ s^{-1}}, \tag{28}$$

and for $N$ lenses the time required for the 2–$\sigma$ upper limit ($\Delta\chi^2 = 4$) to reach the level of $v_{0,rms}$ is

$$T = 10 \left[\frac{v_{0,rms}}{600 \text{ km s}^{-1}}\right] \left[\frac{\Delta\chi^2}{4N}\right]^{1/2} \left[\frac{\gamma}{0.1}\right] \left[\frac{\sigma_\theta}{10\mu as}\right] \left[\frac{D_{OS}}{4g \text{ Gpc}}\right] \text{yrs.} \quad (29)$$

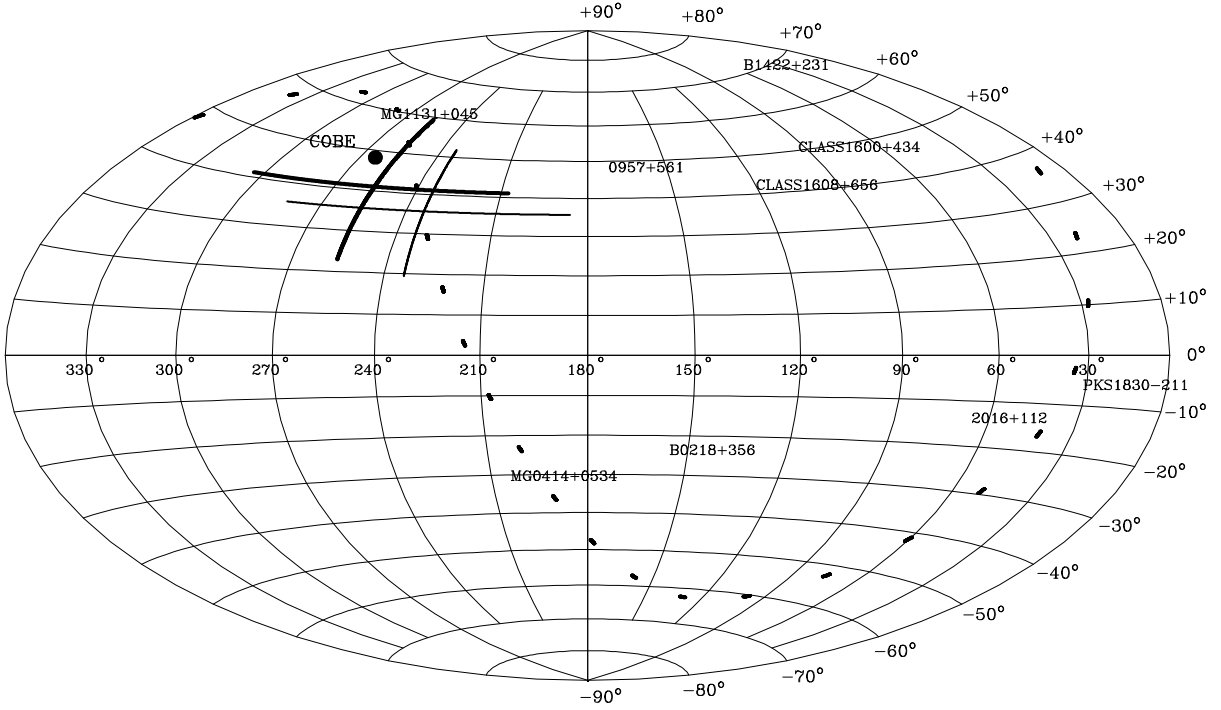

Fig. 3.— Predicted inertial frame dipole direction (bold cross) as calculated by 50 Monte-Carlo runs on known VLBI sources locations with varying velocities and measurement errors. All lenses are considered to be at $z = 0.3$. The heliocentric CMB dipole is marked by the filled dot. A thinner cross is plotted for comparison, when we make all the lenses rotate about the Galactic pole with $\langle \omega_l^2(z) \rangle^{1/2}$. The figure is in Galactic coordinates. The dotted line marks the celestial equator.

Figure 4 shows the expected rms proper motions over ten years in a specific, simplified, family of large-scale structure models. The lenses are all at a redshift of $z_l = 0.3$, and the proper motions were all magnified by the typical value of 34. We used a COBE normalized, CDM power spectrum for open cosmologies ("OCDM" models with $\Lambda_0 = 0$) and assumed that the lenses were a fair sample of galaxies. Models with small Hubble constants and low matter densities have very small rms proper motions (an rms of 1.4 $\mu as$ after ten years for $\Omega_0 = 0.2$, $H_0 = 40$ km s$^{-1}$ Mpc$^{-1}$) while models with high Hubble constants and high matter densities show large proper motions (an rms of 150 $\mu as$ after ten years for $\Omega_0 = 1.0$, $H_0 = 80$ km s$^{-1}$ Mpc$^{-1}$). Small Hubble constants reduce the proper motions because the distances to the lenses are larger, while low matter densities produce smaller rms peculiar velocities. Similar result holds for a flat universe model ($\Lambda$CDM) with $\Omega_0$ and $H_0$ as variables. In these models we assume the theoretically calculated rms peculiar velocities from eq. (2), not the observational estimates of the rms peculiar velocity locally.

We do not escape systematic problems by using lens peculiar velocities. In particular, the lenses are a biased sample, because most lenses [roughly 7 out of 8, Kochanek (1996)] are formed by E and S0 galaxies rather than spirals – the higher masses of the E/S0 galaxies more than make up for their lower number density compared to spiral galaxies. The morphology-density relation (Postman & Geller 1984, Dressler 1980) shows that E/S0 galaxies are ten times more likely to be in or near clusters of galaxies, so the sample defined by the lenses is biased to represent the pattern of velocities in denser regions. Since the pair-wise velocity dispersion of galaxies increases with the density of the sample environment (Marzke 1995), the rms peculiar velocity of lens galaxies will be higher than average. This occurs even though the higher density environment has only weak effects on the lensing properties of the galaxy – we know from the morphology of the lenses that the lens galaxies do not lie in the inner $\sim 1$ $h_{75}^{-1}$Mpc of any cluster. However, the velocities will be strongly modified if the galaxy lies within 3-5 $h_{75}^{-1}$Mpc of the cluster center.[3] Thus the peculiar velocities of lens galaxies will typically be higher than the average peculiar velocities of randomly selected galaxies or field galaxies. This makes measuring the proper motions easier, but their interpretation more difficult.

---

[3]The perturbation a cluster causes on the lens model depends on the projected gravitational acceleration produced by the cluster at the radius of the lens galaxy and is related to the weak shear of the cluster. Outside the cluster core, this is only a small perturbation on the lens model. The perturbation in the velocity of the lens galaxy by the cluster is much larger because the velocity perturbation is the time integral of the weak acceleration.

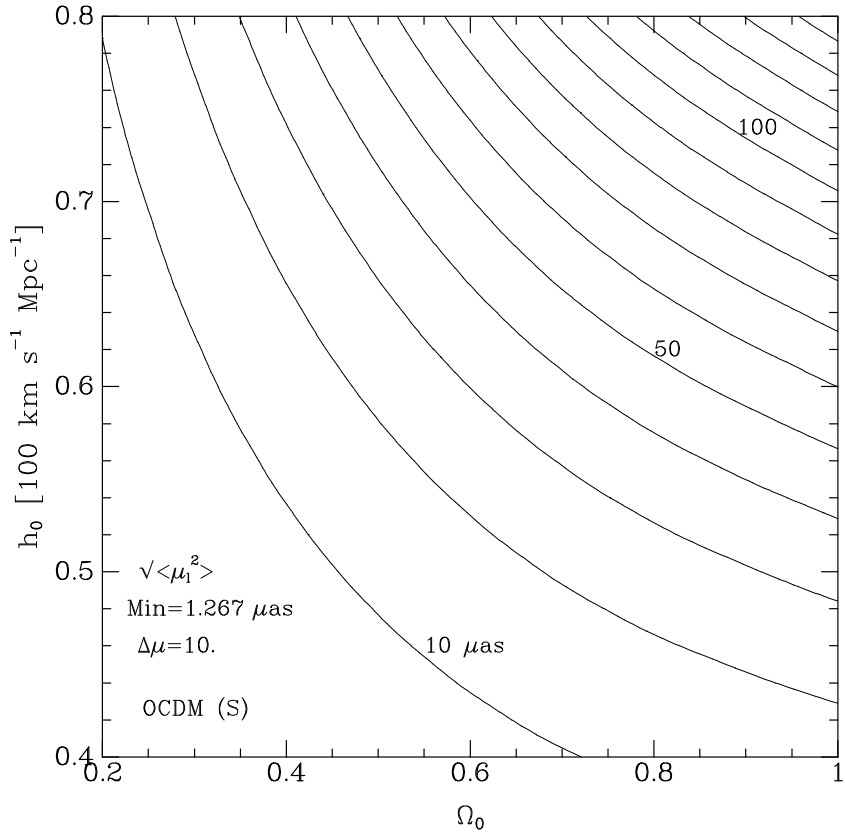

Fig. 4.— Expected *measured* proper motion deviations about the mean (dipole). Velocities and proper motion distances are for $z_l = 0.3$ in an open universe CDM (OCDM) model with the average magnification of $\sim 34$. Contours are spaced from the minimum at the lower left corner (1.27$\mu$as) by $\Delta\mu = 10\mu$as .

## 7. Other Sources of Proper Motions

Peculiar velocities are not the only possible source of proper motions in lensed VLBI sources. Proper motions may also be produced by motion in the source AGN, and possibly by microlensing in the lens galaxy. Although the peculiar velocity of the galaxy containing the AGN is negligible, the velocity of the source may be dominated by internal motions near the AGN. If typical sources have effective internal three-dimensional velocities $v_s$, then the internal motions dominate the proper motion residuals in an $\Omega_0 = 1$ model if $v_s \gtrsim (15g)^{1/2}\omega_{0,rms} = 3.9(1+z_s)^{1/9}\omega_{0,rms}$. If we take $\omega_{0,rms} = \sqrt{2/3}\ 600 = 490$ km s$^{-1}$ then source motions dominate if they are larger than $v_s \gtrsim 1700$ km s$^{-1}$.

The only effect that mimics peculiar velocities over long time scales is motion of the supermassive black hole powering the AGN relative to the source galaxy. Objects in orbits smaller than $3 \times 10^{17}$ cm at a distance of 4 Gpc have maximum proper motions smaller than 10 $\mu$as and have no effect. Orbits larger than $10^{19}M_8$ cm, where the mass producing the orbit is $M = 10^8 M_8 M_\odot$, have source velocities that are only 10% corrections to the expected peculiar motions. Between these two limits, only the wider orbits are physically plausible either because of limits from gravitational radiation for black hole binaries, or because of limits on plausible stellar densities for a black hole orbiting in a central star cluster. The maximum plausible orbital effects are comparable to the effects of the lens peculiar velocity, but in most cases the contamination of the peculiar velocity from motion of the black hole powering the AGN should be negligible.

Apparent peculiar velocities can also be produced by the motion of the emitting material relative to the black hole. The motion may either be orbital motion or non-gravitational motions produced in the jet. Radio emission from orbiting material is unlikely to be important. If we suppose that the emission is created by $N$ patches with velocities generated by the gravity of the black hole, then the maximum characteristic velocity dispersion of the mean centroid of the emission is of order

$$\sigma_s \sim 3000 N^{-1/2} M_8^{1/2} D_{Gpc}^{-1/2} \left(\frac{10\mu as}{\sigma_s}\right)^{1/2} \text{ km s}^{-1} \tag{30}$$

where $\sigma_s$ is the angular size of the emission region. Unless $N \sim 1$, we do not expect the centroid of the radio emission region to show a large enough velocity relative to the AGN to be important, and there is no reason for orbiting material to reach this limit.

The remaining case, and the one known to be important from the existence of superluminal sources, is motion of material in the AGN jet. The effects of source motions and the possible relations between gravitational lensing and normal superluminal motions are discussed by Chitre & Narlikar (1979, 1980, also Chitre & Saslaw 1989) and Gopal-Krishna & Subramanian (1991, 1996). For our purposes, we need not worry about classical superluminal motion. A lensed classical superluminal source would be easily detected, since the magnified motion of the superluminal components could be tens of *milliarcseconds* per year. Suppose the core AGN has flux $f_1$ and a blob is emitted with flux $f_2$ and apparent transverse velocity of $\omega$. Then the apparent motion of an unresolved source is $\omega f_2/(f_1 + f_2)$. Sufficiently fast and bright components are detectable as normal superluminal motion, so the primary problems are superluminal but very low flux components, and "sub-luminal" bright components. Since such motions are, by definition, not directly observed, the only way to eliminate them as a source of error is to monitor the source for extended periods of time – the true proper motions will continue to grow, while internal source motions will be an extra source of stochastic noise. If the motions seen in 1038+258 A/B or 0957+561 discussed in §4 are real rather than systematic errors, then proper motions will be trivially detected in lenses. A quad lens should magnify the $10 - 20\mu$as yr$^{-1}$ motions to $300 - 600\mu$as yr$^{-1}$.

The other process that may affect the proper motion measurement is microlensing of the source by the granularity in the potential of the lens. Gopal-Krishna & Subramanian (1991, 1996) have suggested that microlensing may produce observable effects in compact double radio sources. The overall magnification of the source is produced by a combination of the smooth potential creating the multiple images, and microlensing by the fluctuations in the gravitational field from the stars in the lens galaxy. The effects of motion have mainly been discussed in the context of microlensing the optical cores of AGN (Kundić & Wambsganss 1993; Wambsganss & Kundić 1995; Kundić, Witt & Chang 1993). While microlensing is detected in several optical lenses as time variations of the optical flux (e.g., Irwin et al. 1989; Corrigan et al. 1991; Racine 1992), it is impossible to determine the velocity of the images from the light curve with any precision even with perfect sampling. The flux variations are a combination of the velocity, the mass spectrum of the lenses, the size of the source, the surface matter density and the shear. It is also a stochastic process, so unlike the proper motions of the primary images there is no signal that monotonically increases with time. Microlensing by solar-mass stars acts on a characteristic angular scale of $10^{-6}$ arcsec while the size of typical VLBI sources is of order $10^{-3}$ arcsec. Hence microlensing by individual stars affects a fraction of $10^{-6}$ of the source size and produces a negligible effect on the total flux of the sources. Although the combined effect of an ensemble of stars can affect a much larger region of the source (typically about ten times larger, Katz, Balbus, & Paczyński 1986), the magnification of the total flux by microlensing remains negligible. In this case, microlensing cannot induce any detectable motions in the images. For the characteristic microlensing scale to be of order $10^{-3}$ arcsec, the microlenses would have to have $\sim 10^6\,M_\odot$, typical for globular clusters, but the optical depth for microlensing by globular clusters is very small. Even then, the random motion of the microlenses within the lensing galaxy is of order 100 km s$^{-1}$ or about an order of magnitude smaller than the transverse velocity of the macrolens, rendering possible microlensing-induced image velocities very small. Although the velocity of microlensing caustics can be much larger than that of the microlenses, random velocities of microlenses in the lensing galaxy are on average only $\sim 25\%$ more efficient in producing time variable magnifications of the source than the transverse velocity of the galaxy as a whole (Kundić & Wambsganss 1993). Finally, microlensing would only affect the flux of one source image at a time so that the effect of microlensing could be identified by using the relative velocities of all source images.

## 8. Conclusion and Discussion

Detectable proper motions in the quad lenses observable with VLBI are inevitable, and the only issues are the time scale required to detect the motions, and whether the motions will be dominated by the peculiar velocities of the observer and the lens galaxy or motions in the source AGN itself. Although we focused on the study of peculiar velocities, either result is physically interesting. Useful limits can be set given temporal baselines of 10 years, and it is actuarially certain that proper motions will be measured in all of the VLBI quads before the authors of the paper retire. For the moment, the only observational requirement is that VLBI observations of the quad lenses should be made keeping the goal of determining accurate proper motions in mind. The proper motions expected for double lenses are an order of magnitude smaller than in quad lenses, so it comes as no surprise that there is no convincing determination of a relative proper motion in the 0957+561 lens (Campbell 1995, Campbell et al. 1994).

The proper motions are dominated by the proper motions of the lens galaxies, because most estimates of the rms peculiar velocities of galaxies are significantly larger than our motion relative to the CMB frame. If the rms peculiar velocity today is $v_{rms,0}$ and we assume it evolves with redshift as $v_{rms}(z) = v_{rms,0}/(1+z)^{1/2}$, then the time scale for a 2–$\sigma$ detection in a typical quad lens is

$$T = 10 \left[ \frac{v_{0,rms}}{600 \text{ km s}^{-1}} \right] \left[ \frac{\sigma_\theta}{10\mu as} \right] \left[ \frac{D_{OS}}{4g(z_s) \text{ Gpc}} \right] \text{yrs}, \qquad (31)$$

assuming that the rms uncertainty in the measurement of one coordinate of an image position is $\sigma_\theta = 10$ $\mu$as (optimistic), but that only two measurements are made, and using only one image pair instead of all six independent measurements (pessimistic). Moreover, our statistical model uses an underestimate of the rms magnification of the proper motions.

Given an ensemble of quad lenses with proper motions we can estimate the proper motion dipole due to our motion relative to the extragalactic inertial frame. In any one system, the typical magnitude of the proper motion dipole is smaller than the peculiar dipole, but it is detectable when averaged over an ensemble of lenses. Given ten lenses and an rms uncertainty of $\sigma_\theta = 10$ $\mu$as, we estimate that the direction of the dipole can be determined to 20° and the Hubble constant can be determined to $\Delta H_0/H_0 = 0.34$ after ten years (1–$\sigma$ limits). After ten years the uncertainties are dominated by the motions of the lens and the source rather than measurement error, so precision measurements depend on finding larger numbers of lenses.

Barrow, Juszkiewicz, and Sonoda (1985) put limits on the dimensionless quantity $\omega/H_0$ by using constraints from the quadrupole moment in the CMB multipole expansion. These limits are still relevant in view of the current COBE results (Bennett et al. 1994) of $\omega/H_0 < 2 \times 10^{-5}$ $(10^{-4})$ for a flat (open) universe. In turn, these numbers permit a maximum rotation rate of $10^{-14}\mathrm{rad}\,\mathrm{yr}^{-1} \simeq 0.02$ $\mu$as yr$^{-1}$. The rotation limits of the "universe" (i.e. our horizon) from CMB quadrupole ($\omega \lesssim 0.02\mu\mathrm{as}\,\mathrm{yr}^{-1}$) are stronger than the possible limits from the current proposed analysis. Rotation, however, can also be contributed from a broken anisotropy on smaller scales than the horizon. In the latter case the current method can put limits of the order $\omega < 10^{-11}\,\mathrm{rad}\,\mathrm{yr}^{-1}$.

The systematic uncertainties in this approach are dominated by the poor limits on the internal motions of VLBI sources that are not obvious superluminals. If these motions exceed $\sim 2000$ km s$^{-1}$, they will dominate the proper motions on short time scales. Internal motions and microlensing do not affect the longer term proper motion of the lens, because both processes only introduce extra noise in addition to the monotonic proper motions produced by the velocity of the observer, lens, and source galaxies. Existing estimates for proper motions of extragalactic VLBI sources are consistent with zero given their systematic errors (e.g. Marcaide et al. (1994), Campbell (1995), Campbell et al. (1994)). If, however, typical sources had apparent motions even 10% of the values seen in 1038+528 A/B ($31 \pm 22$ $\mu$as yr$^{-1}$) or 0957+561 ($\sim 25$ $\mu$as yr$^{-1}$) under the assumption that the motion is real, then the VLBI quad lenses should show astounding proper motions ($\sim 300$ $\mu$as yr$^{-1}$). While we view this as unlikely, such proper motions would be extraordinarily useful for constraining the lens geometry and probing AGN physics.

We would like to thank L. Greenhill, J.N. Hewitt, J.M. Moran, and I.I. Shapiro for discussions, and J. Miralda-Escudé and U. Seljak for pointing out an error. This work was supported in part by NSF grants PHY-91-06678 (TK) and AST-9401722 (CSK).

---

This preprint was prepared with the AAS LaTeX macros v4.0.